\documentclass[aps,prd,preprintnumbers,nofootinbib,twocolumn]{revtex4}
\usepackage{bm}
\usepackage{latexsym}
\usepackage{dcolumn}
\usepackage{amsmath,amsfonts,amssymb}
\usepackage{graphicx,epsfig}
\usepackage{psfrag}
\usepackage{amsthm}

\def\be {\begin{equation}}
\def\ee {\end{equation}}
\def\bea {\begin{eqnarray}}
\def\eea {\end{eqnarray}}
\def\bc {\begin{center}}
\def\ec {\end{center}}
\def\bfg {\begin{figure}}
\def\efg {\end{figure}}
\def\bi {\begin{itemize}}
\def\ei {\end{itemize}}
\def\nn {\nonumber}

\def\la {\label}
\def\le {\left}
\def\ri {\right}
\def\pa {\partial}

\def\vs {\vspace}

\def\a  {\alpha}
\def\b  {\beta}

\def\d  {\delta}
\def\D  {\Delta}

\def\s {\sigma}

\def\beq{\begin{equation}}
\def\eeq{\end{equation}}
\def\br{\begin{eqnarray}}
\def\er{\end{eqnarray}}
\newcommand{\eel}[1] {\label{#1}\end{equation}}

\newcommand{\bdm}{\begin{displaymath}}
\newcommand{\edm}{\end{displaymath}}

\begin{document}
\title{Discreteness of Space from GUP II: Relativistic Wave Equations}

\author{Saurya Das $^1$} \email[email: ]{saurya.das@uleth.ca}
\author{Elias C. Vagenas $^2$} \email[email: ]{evagenas@academyofathens.gr}
\author{Ahmed Farag Ali $^1$} \email[email: ]{ahmed.ali@uleth.ca}

\affiliation{$^1$~Dept. of Physics,
University of Lethbridge, 4401 University Drive,
Lethbridge, Alberta, Canada T1K 3M4 \\}

\affiliation{$^2$~Research Center for Astronomy \& Applied Mathematics,\\
Academy of Athens, \\
Soranou Efessiou 4,
GR-11527, Athens, Greece
}

\begin{abstract}
Various theories of Quantum Gravity predict modifications of the Heisenberg Uncertainty Principle near
the Planck scale to a so-called Generalized Uncertainty Principle (GUP). In some recent papers, we
showed that the GUP gives rise to corrections to the Schrödinger equation, which in turn affect all
quantum mechanical Hamiltonians. In particular, by applying it to a particle in a one-dimensional box,
we showed that the box length must be quantized in terms of a fundamental length (which could be the
Planck length), which we interpreted as a signal of fundamental discreteness of space itself. In this Letter,
we extend the above results to a relativistic particle in a rectangular as well as a spherical box, by solving
the GUP-corrected Klein–Gordon and Dirac equations, and for the latter, to two and three dimensions. We
again arrive at quantization of box length, area and volume and an indication of the fundamentally grainy
nature of space. We discuss possible implications.
\end{abstract}

\maketitle


Various approaches to quantum gravity
(such as String Theory and Doubly Special Relativity (or DSR) Theories),
as well as black hole physics, predict a minimum measurable length, and a
modification of the Heisenberg Uncertainty Principle to a so-called
Generalized Uncertainty Principle, or GUP, and a corresponding modification of the
commutation relations between position coordinates
and momenta. The following GUP which we proposed in \cite{advplb}
is (and as far as we know the only one) consistent with DSR theories, String Theory and
Black Holes Physics {\it and}
which ensure $[x_i,x_j]=0=[p_i,p_j]$ (via the Jacobi identity)
\footnote{
(a) In \cite{advplb,dvprl,dvcjp} we had used $\a$ in place of $a$.  \\
(b) The results of this article
do not depend on this particular form of GUP chosen,
and continue to hold for a large class of variants,
so long as an ${\cal O}(a)$ term is present in the right-hand side
of Eq.(\ref{comm01}).
}
\bea
[x_i, p_j]\hspace{-1ex} &=&\hspace{-1ex} i \hbar\hspace{-0.5ex} \left[  \delta_{ij}\hspace{-0.5ex}
- \hspace{-0.5ex} a \hspace{-0.5ex}  \le( p \delta_{ij} +
\frac{p_i p_j}{p} \ri)
+ a^2 \hspace{-0.5ex}
\le( p^2 \delta_{ij}  + 3 p_{i} p_{j} \ri) \hspace{-0.5ex} \ri]
\label{comm01} \\
%
%
%
 \Delta x \D p \hspace{-1ex}&\geq &\hspace{-1ex}\frac{\hbar}{2}
\le[ 1 - 2 a <p> + 4a^2 <p^2>
\ri] ~ \nn \\
\hspace{-1ex}&\geq& \hspace{-1ex}
\frac{\hbar}{2} \hspace{-1ex}
\le[\hspace{-0.5ex} 1\hspace{-0.5ex}  +\hspace{-0.7ex}  \le(\hspace{-0.7ex}
\frac{a}{\sqrt{\langle p^2 \rangle}} +\hspace{-0.2ex}4a^2 \hspace{-0.9ex} \ri)
\hspace{-0.6ex}  \D p^2 \hspace{-0.6ex}
+\hspace{-0.6ex}  4a^2 \langle p \rangle^2 \hspace{-0.6ex}
- \hspace{-0.6ex}  2a \sqrt{\langle p^2 \rangle}\hspace{-0.2ex}
\ri]\hspace{2ex} \label{dxdp1}
\eea
where
$a = {a_0}/{M_{Pl}c} = {a_0 \ell_{Pl}}/{\hbar},$
$M_{Pl}=$ Planck mass, $\ell_{Pl}\approx 10^{-35}~m=$ Planck length,
and $M_{Pl} c^2=$ Planck energy $\approx 10^{19}~GeV$.
It should be stressed that the GUP-induced terms become important near the Planck scale.
It is normally assumed that $a_0 \approx 1$.
(For earlier versions of GUP, motivated by String Theory, Black
Hole Physics, DSR etc, see e.g.
\cite{guppapers,kmm,kempf,brau,sm,cg}, and for some
phenomenological implications see \cite{dvprl,dvcjp,advplb}.) Note
that although Eqs. (\ref{comm01}) and (\ref{dxdp1}) are not
Lorentz covariant, they are at least approximately covariant under
DSR transformations \cite{cg}. We expect the results of our Letter
to have similar covariance as well. In addition, since DSR transformations
preserve not only the speed of light, but also the Planck momentum
and the Planck length, it is not surprising that Eqs.
(\ref{comm01}) and (\ref{dxdp1}) imply the following minimum
measurable length {\it and} maximum measurable momentum
\bea
\D x &\geq& (\D x)_{min}  \approx a_0\ell_{Pl} \la{dxmin} \\
\D p &\leq& (\D p)_{max} \approx \frac{M_{Pl}c}{a_0}~. \la{dpmax}
\eea
\par\noindent
It can be shown that the following definitions
\bea x_i = x_{0i}~,~~
p_i = p_{0i} \le( 1 - a p_0 + 2a^2 p_0^2 \ri)~, \la{mom1}
\eea
(with $x_{0i}, p_{0j}$
satisfying the canonical commutation relations
$ [x_{0i}, p_{0j}] = i \hbar~\delta_{ij}, $
such that $p_{0i} = -i\hbar \partial/\partial{x_{0i}}$)
satisfy Eq.(\ref{comm01}).
In \cite{advplb} we showed that any non-relativistic Hamiltonian of the form
$H=p^2/2m + V(\vec r)$ can be written as $H = p_0^2/2m - (a/m)p_0^3 + + V(r) + {\cal O}(a^2)$ using
Eq.(\ref{mom1}), where the second term can be treated as a perturbation. Now, the
third order Schr\"odinger equation has a new {\it non-perturbative} solution of the form
$\psi \sim e^{ix/2a\hbar}$, which when superposed with the regular solutions perturbed
by terms ${\cal O}(a)$, implies not only the usual quantization of energy, but also that the
box length $L$ is quantized according to
\bea
\frac{L}{a\hbar} = \frac{L}{a_0 \ell_{Pl}} = 2p\pi + \theta~,~p \in \mathbb{N}
\la{quant1}
\eea
where $\theta = {\cal O}(1)$. We interpreted this as the quantization of measurable lengths, and
effectively that of space itself, near the Planck scale. In this Letter, we re-examine the above problem,
but now assuming that the particle is relativistic. This we believe is important for several reasons, among which are that extreme high energy (ultra)-relativistic particles are natural candidates for probing the nature of spacetime near the Planck scale, and that most elementary particles in nature are fermions, obeying
some form of the
Dirac equation. Furthermore, as seen from below, attempts to extend our results to
$2$ and $3$ dimensions seem to necessitate the use of matrices.
However, we first start by examining the simpler Klein-Gordon equation.


\section{Klein-Gordon Equation in One Dimension}

The Klein-Gordon (KG) equation in $1$-spatial dimension
\footnote{
In this and in subsequent sections, we start with the usual forms of the KG and Dirac
equations, as is indeed the case for massless particles, as well as
for massive particles {\it and} in their stationary states, with the re-definition
$m \rightarrow m (1-\ell_{Pl} E/\hbar c)$, see e.g. Eq.(11) of \cite{sm} or Eq.(14) of \cite{ghoshetal}.
}
\bea
%
 p^2 \Phi(t,x) = \le(  \frac{E^2}{c^2} - m^2c^2 \ri) \Phi (t,x)~.
\la{kg1}
%
%
\la{kg2}
\eea
%
%
%
%
%
We see that this is identical to the Schr\"odinger equation, when one makes the
identification:
$
2mE/\hbar^2 \equiv k^2 \rightarrow E^2/\hbar^2 c^2 - m^2 c^2/\hbar^2~.
$
As a result, the quantization of length, which does not depend
on $k$, continues to hold \cite{advplb}.

However, in addition to fermions being the most fundamental entities,
the $3$-dimensional version of
KG equation (\ref{kg1}), when combined with Eq.(\ref{mom1}),
suffers from the drawback that the $p^2$ term translates to
$p^2 = p_0^2 - 2a p_0^3 + {\cal O}(a^2)
= - \hbar^2 \nabla^2 +i 2a \hbar^3 \nabla^{3/2} + {\cal O}(a^2)$,
of which the second term is evidently non-local.
As we shall see in the next section,
the Dirac equation can address both issues at once.

%

\section{Dirac Equation in One Dimension}

First
we linearize
$p_0 = \sqrt{p_{0x}^2+p_{0y}^2+p_{0z}^2}$ using the Dirac prescription, i.e. replace
$p_0 \rightarrow \vec\alpha \cdot \vec p$,
where $\alpha_i~(i=1,2,3)$ and $\b$ are the Dirac matrices, for which we
use the following representation
\bea
\a_i =
\left( \begin{array}{cc}
0 & \sigma_i \\
\sigma_i & 0 \end{array} \right)~,~
\b =
\left( \begin{array}{cc}
I & 0 \\
0 & -I \end{array} \right)~.
\eea
The GUP-corrected Dirac equation can thus be written to ${\cal O}(a)$ as
\footnote{In this section, we closely follow the formulation of \cite{afg}.}
\bea
H \psi &=& \le (c\, \vec \a \cdot \vec p + \b mc^2 \ri) \psi (\vec r)  \nn \\
&=&  \le(c\, \vec \a \cdot \vec p_0 -
c\, a (\vec\a \cdot \vec p_0)(\vec\a \cdot \vec p_0) + \b mc^2 \ri) \psi (\vec r)\nn \\
&=& E\psi (\vec r)
\la{ham1}
\eea
which for $1$-spatial dimension, say $z$, is in the position representation
\bea
\le(
-i\hbar c \a_z \frac{d}{dz} + c a\hbar^2 \frac{d^2}{dz^2} + \b mc^2
\ri)\psi(z) = E\psi (z)~.
\la{diraceqn1}
\eea
Note that this is a second order differential equation instead of the usual
first order Dirac equation
(we have used $\a_z^2=1$). Thus, it has two linearly independent, positive energy
solutions, which to ${\cal O}(a) $ are
\bea
\psi_1 &=& N_1~e^{ikz} \left( \begin{array}{c}
\chi \\
{} \\
r \sigma_z \chi \end{array} \right) \\
{} \nn \\
\psi_2 &=& N_2~e^{i\frac{z}{a\hbar}} \left( \begin{array}{c}
\chi \\
{} \\
\sigma_z \chi \end{array} \right)
\eea
where $m$ is the mass of the Dirac particle, $k =k_0 + a\hbar k_0^2$,
$k_0$ satisfies the usual dispersion relation $E^2 = (\hbar k_0 c)^2 + (mc^2)^2$,
$r \equiv \frac{\hbar k_0 c}{E+mc^2}$ and
$\chi^\dagger \chi=I$.
Note that $r$ runs from $0$ (non-relativistic) to $1$ (ultra-relativistic).
$k,k_0$ could be positive (right moving) or negative (left moving).
$N_1,N_2$ are suitable normalization constants.
As in the case of Schr\"odinger equation, here too
a new {\it non-perturbative} solution $\psi_2$ appears, which
should drop out in the $a \rightarrow 0$ (i.e no GUP) limit. This has a characteristic
wavelength $2\pi a \hbar$.

As noted in \cite{afg}, to confine a relativistic particle in a box of length $L$
in a consistent way avoiding the Klein paradox (in which an increasing number of
negative energy particles are excited), one may take its
mass to be $z$-dependent as was done in
the MIT bag model of quark confinement
\bea
m(z) &=& M,~z<0~~~\mbox{(Region I)} \nn\\
     &=& m,~ 0 \leq z \leq L~~~\mbox{ (Region II)}\nn \\
     &=& M,~z>L~~~\mbox{ (Region III)},
\eea
where $m$ and $M$ are constants and we will eventually
take the limit $M \rightarrow \infty$.
Thus, we can write the general wavefunctions in the three regions
\bea
\psi_I &=& A~e^{-iKz}
\left( \begin{array}{c}
\chi \\
{} \\
-R \sigma_z \chi
\end{array} \right)
%
+
G~e^{i\frac{z}{a\hbar}} \left( \begin{array}{c}
\chi \\
{} \\
\sigma_z \chi \end{array} \right) \la{psi1} \\
%
%
\psi_{II} &=& B~e^{ikz}
\left( \begin{array}{c}
\chi \\
{} \\
r\sigma_z \chi
\end{array} \right)
+ C~e^{-ikz}
\left( \begin{array}{c}
\chi \\
{} \\
-r\sigma_z \chi
\end{array} \right)
\nn \\
&+&
F~e^{i\frac{z}{a\hbar}} \left( \begin{array}{c}
\chi \\
{} \\
\sigma_z \chi \end{array} \right)
\la{psi2} \\
%
%
\psi_{III} &=& D~e^{iKz}
\left( \begin{array}{c}
\chi \\
{} \\
R \sigma_z \chi
\end{array} \right)
%
+
H~e^{i\frac{z}{a\hbar}} \left( \begin{array}{c}
\chi \\
{} \\
\sigma_z \chi \end{array} \right)~, \la{psi3}
\eea
where $E^2 = (\hbar K_0 c)^2 + (Mc^2)^2$, $K =K_0 + a\hbar K_0^2$ and $R=\hbar K_0c/(E+Mc^2)$.
Thus, in the limit $M\rightarrow \infty$, $K \rightarrow +i\infty$,
the terms associated with $A$ and $D$ go to zero. However, those with $G$ and
$H$ do not. Moreover, it can be shown that
the fluxes due to these terms do not vanish. Thus, we must set $G=0=H$.
In addition, without loss of generality we choose $B=1$ and $C=e^{i\delta}$ where $\delta$ is
a real number. It can be shown that if one chooses $|C| \neq 1$ then the energy
of the relativistic particle is complex.
Finally, we must have $F \sim a^s~,s>0$, such that this term goes to zero in the
$a \rightarrow 0$ limit.
Now, boundary conditions akin to that for the Schr\"odinger equation, namely
$\psi_{II}=0$ at $z=0$ and $z=L$ will require $\psi_{II}$ to vanish identically.
Thus, they are disallowed. Instead, we require
the outward component of the Dirac current to be zero at the boundaries (the MIT bag model).
This ensures that the particle is indeed confined within the box \cite{thomas}.

The conserved current corresponding to Eq.(\ref{diraceqn1}) can be shown to be
\bea
J_z &=& \bar\psi \gamma^z \psi + ic\hbar a \le(
\psi^\dagger \frac{d\psi}{dz} - \frac{d\psi^\dagger}{dz} \psi
\ri)~, \nn \\
&\equiv& J_{0z} + J_{1z} \la{current1}
\eea
where $J_{0z} +J_{1z}$ are the usual and new GUP-induced currents, respectively.
We will comment on $J_{1z}$ shortly.
First, the vanishing of the Dirac current $J^\mu = \bar\psi\gamma^\mu\psi$ at a
boundary is equivalent to the condition $i\gamma\cdot n \psi = \psi$ there,
where $n$ is the outward normal to the boundary \cite{thomas}.
Applying this to $J_{0z}$ for the wavefunction $\psi_{II}$ at $z=0$ and $z=L$ gives
\cite{afg}
\bea
i\b\a_z\psi_{II}\big|_{z=0} &=& \psi_{II}\big|_{z=0}  \la{mitbc1} \\
\mbox{and}~
-i\b\a_z\psi_{II}\big|_{z=L} &=& \psi_{II}\big|_{z=L} \la{mitbc2}~,
\eea
respectively. Using the expression for $\psi_{II}$ from (\ref{psi2}), we get
from (\ref{mitbc1}) and (\ref{mitbc2}), respectively,
\bea
\frac{B+C + F' e^{-i\pi/4}}{B-C} &=& ir \\
\frac{B e^{ikL} + Ce^{-ikL} + F' e^{i(L/a\hbar + \pi/4})}{Be^{ikL}-Ce^{-ikL}}
&=& -ir ~
\eea
(where $F'= \sqrt{2} F$), which in turn yield
\bea
(i r -1) - F'e^{-i \pi/4} = (ir+1) e^{i\d} \hspace{1ex}\la{cond3}&& \\
(ir -1) - F'e^{i(L/a\hbar + \pi/4)} e^{ikL} e^{-i \delta}=&&\nn\\
(ir+1) e^{i(2kL-\delta)}~.&&\hspace{1ex} \la{cond4}
\eea
%
Note that conditions (\ref{cond3}) and (\ref{cond4}) imply
\bea
|B| = |C| + {\cal O}(a)~,
\eea
which guarantees that
\bea
J_{1z} = -2 ca\hbar k (1+r^2) \le[ |B|^2 - |C|^2\ri] =0~.
\eea
Furthermore, from (\ref{cond3}) and (\ref{cond4}) it follows that
\bea
kL  &=& \d = \arctan \le( -\frac{\hbar k}{mc} \ri) + {\cal O}(a) \la{kl}\\
\mbox{and}~~\frac{L}{a\hbar} &=&\frac{L}{a_0 \ell_{Pl}}
= 2p\pi - \frac{\pi}{2}~,~~p \in \mathbb{N}~. \la{1dquant}
\eea
The transcendental equation (\ref{kl}) gives the quantized energy levels
for a relativistic particle in
a box. Its $a\rightarrow 0$ limit gives $k_0 L= \arctan \le( -\frac{\hbar k_0}{mc} \ri)$
which is Eq.(17) of ref.\cite{afg}, its non-relativistic limit gives $(k_0+a\hbar k_0^2) L=n\pi$, while
its non-relativistic {\it and} $a\rightarrow 0$ limit yields the Schr\"odinger equation result $k_0L=n\pi$.
Equation (\ref{1dquant}) on the other hand shows that such a particle cannot be confined
in a box, unless the box length is quantized according to this condition.
Note that this is {\it identical} to the quantization condition (\ref{quant1}), which was derived using the Schr\"odinger equation
(with the identification $\theta \equiv -\pi/2$). This indicates the robustness of the result.
As measuring spatial dimensions requires the existence and observation of
at least one particle, the above result once again seems to indicate that
effectively all measurable lengths are quantized in units of $a_0\ell_{Pl}$.
%
%
\section{Dirac Equation in Two and Three Dimensions}
%
%
We now generalize to a box in two or three dimensions defined by
$0 \leq x_i \leq L_i,i=1,\dots,d$ with $d=1,2,3$.
We start with the following ansatz for the wavefunction
\bea
\psi =
e^{i \vec t \cdot \vec r}
\left( \begin{array}{c}
\chi \\
{} \\
\vec \rho \cdot \vec \sigma \chi \end{array} \right)
\eea
where $\vec t$ and $\vec \rho$ are d-dimensional (spatial) vectors,
and $\chi^\dagger\chi=I$ as before.
In this case, Eq.(\ref{ham1}) translates to
\bea
H\psi \hspace{-1ex}&=&\hspace{-1ex}
e^{i \vec t \cdot \vec r}
\left( \hspace{-1ex}\begin{array}{c}
( (mc^2 - c a \hbar^2 t^2 ) + c \hbar \left( \vec t \cdot \vec \rho + i \vec \sigma \cdot ({\vec t \times \vec \rho) }\right) ) \chi \\
{} \\
\le( c \hbar \vec t - (m c^2   + c  a \hbar^2 t^2 ) \vec \rho \ri) \cdot \vec \sigma \chi \end{array} \hspace{-1ex}\right) \nn \\
&=& E\psi~,
\la{dirac23}
\eea
where we have used the identity
$(\vec t \cdot \vec \sigma)(\vec \rho \cdot \vec \sigma) =  \vec t \cdot \vec \rho + i \vec \sigma
\cdot (\vec t \times \vec \rho)$.
Eq.(\ref{dirac23}) implies $\vec t \times \vec \rho = 0$, i.e. $\vec \rho$ is parallel to $\vec t$, and
two solutions for $t$, namely $t=k$ and $t=1/a\hbar$, and correspondingly
$\rho = r$ and $\rho = 1$. The latter
solutions for $t$ and $\rho$ are the (new) non-perturbative ones,
which as we shall see, will give rise to quantization of space.
Thus the vector $\vec t$ for the two cases are
$\vec t = \vec k$ and $\vec t = \frac{\hat q}{a\hbar}$
and
$\vec \rho = r \hat k$ and $\vec\rho=\hat q$ respectively,
where $\hat q$ is an arbitrary unit vector
\footnote{Although one can choose $\hat q = \hat k$, {\it per se}
our analysis does not require this to be the case. We will comment on this towards the
end of the Letter.}.
Thus, putting in the normalizations, the two independent positive energy
solutions are


\bea
\psi_1 &=& N_1
e^{i \vec k \cdot \vec r}
\left( \begin{array}{c}
\chi \\
{} \\
r \hat k \cdot \vec \sigma \chi \end{array} \right)  \\
\psi_2 &=& N_2
e^{i \frac{\hat q \cdot \vec r }{a\hbar} }
\left( \begin{array}{c}
\chi \\
{} \\
{\hat q} \cdot \vec \sigma \chi \end{array} \right)
\eea
with $\psi_2$ being the new GUP-induced eigenfunction.

Next, we consider the following wavefunction
\bea
\psi =
\left( \begin{array}{c}
\le[
\prod_{i=1}^d
\le(
e^{i k_i x_i} + e^{-i(k_i x_i - \d_i)} \ri)
+ F e^{i  \frac{\hat q \cdot \vec r}{a\hbar} }
\ri] \chi
\\
{} \\
\sum_{j=1}^d
\le[
\prod_{i=1}^d
\le( e^{i k_i x_i} + (-1)^{\d_{ij}} e^{-i(k_i x_i - \d_i)} \ri)
r \hat k_j
 \ri.
 \\
\le.
+ F e^{ i  \frac{\hat q \cdot \vec r}{a\hbar}  }~\hat{q_j} \ri]
\sigma_j \chi
\end{array}
\right)\,\,
\la{psi3d}
\eea
where $d=1,2,3$, depending on the number of spatial dimensions
and an overall normalization has been set to unity. The number of terms in row I and
row II are $2^{d} + 1$ and $(2^d +1 )\times d$ respectively, i.e. $(3,3), (5,10)$ and $(9,27)$ in $1,2$ and
$3$ dimensions, respectively. It can be easily shown that the above
is a superposition of $F \psi_2$ and the following $2^d$ eigenfunctions,
for all possible combinations with $\epsilon_i~(i=1,\dots,d)$, with $\epsilon_i = \pm 1$
\bea
\Psi &=& e^{i(\sum_{i=1}^d \epsilon_i k_i x_i + \frac{(1-\epsilon_i)}{2}\d_i)}
\left( \begin{array}{c}
\chi \\
{} \\
r \sum_{i=1}^d \epsilon_i \hat k_i \sigma_i \chi
\end{array} \right)
%
\eea
where $\d_i~(i=1,\cdots,d)$ are phases to be determined shortly using boundary conditions.
%
%
%
%
%
%

Again, we impose the MIT bag boundary conditions
$\pm i \b \a_k \psi = \psi~,k=1,\cdots,d$, with the $+$ and $-$ signs
corresponding to $x_k=0$ and $x_k=L_k$ respectively, ensuring vanishing
flux through all six boundaries. First, we write the above
boundary condition for any $x_k$, for the wavefunction given in Eq.(\ref{psi3d}).
This yields the following $2$-component equation
\bea
\hspace{-0.3cm}
\pm
\hspace{-0.1cm}
\left( \begin{array}{c}
i
\sum_{j=1}^d
\le[
\prod_{i=1}^d
\le(
e^{i k_i x_i} + (-1)^{\d_{ij}} e^{-i(k_i x_i - \d_i )}\ri) \s_k~ r_j \s_j

\ri.
\\
\le.
+ ~F~e^{i\frac{\hat q \cdot \vec r}{a\hbar}}~\sigma_k~\hat q_j \sigma_j
\ri] \chi
\\
{}\\
-i \le[ \prod_{i=1}^d
\le(
e^{i k_i x_i} + e^{-i(k_i x_i - \d_i)}
\ri)
+F~e^{i\frac{\hat q \cdot \vec r}{a\hbar}}
\ri] \s_k \chi
\end{array}
\right)\hspace{-0.5cm} \nn \\
= \psi ~.~~~~~~~~~~~~~~~~~~~~~~~~~~~~~~~~~~~~~~~~~~~~~~~~~~~~~~~~~~~~~~~~~~~
\label{3dbc}
\eea
Employing the MIT bag model boundary conditions and thus
equating the rows I and II of Eq. (\ref{psi3d}) with the corresponding ones of Eq.(\ref{3dbc})
yields, respectively
\bea
&& \prod_{i=1}^d
\le(
e^{ik_i x_i} + e^{-i(k_i x_i - \d_i)}
\ri)
+ F~e^{i \frac{\hat q \cdot \vec r}{a\hbar}}
\nn \\
&& =
\pm
\Bigg [ i \prod_{i=1}^d
\le(
e^{i k_i x_i} + (-1)^{\d_{ik}}  e^{-i(k_i x_i -\d_i)}
\ri) r \hat k_k  \nn \\
&&
+~i F~e^{i \frac{\hat q \cdot \vec r}{a}}
~\hat q_k \nn \\
&& + ~i \sum_{j=1\neq k}^d
\Big[ \prod_{i=1}^d
\le(
e^{i k_i x_i} + (-1)^{\d_{ij}} e^{-i(k_i x_i -\d_i)}
\ri) r \hat k_j \s_k \s_j  \nn \\
&&  + ~F~e^{i \frac{\hat q \cdot \vec r}{a}}
~\hat q_j \s_k \s_j \Big ] \Bigg ]
\la{row1}
\eea
and
\bea
&& \prod_{i=1}^d
\le(
e^{ik_i x_i} + e^{-i(k_i x_i - \d_i)}
\ri)
+ F~e^{i \frac{\hat q \cdot \vec r}{a\hbar}}
\nn \\
&& =
\pm \Bigg [
i \prod_{i=1}^d
\le(
e^{i k_i x_i} + (-1)^{\d_{ik}}  e^{-i(k_i x_i -\d_i)}
\ri) r \hat k_k  \nn \\
&&
 +~i~F~e^{i \frac{\hat q \cdot \vec r}{a\hbar}}
~\hat q_k \nn \\
&& + ~i \sum_{j=1\neq k}^d \Big [ \prod_{i=1}^d
\le(
e^{i k_i x_i} + (-1)^{\d_{ij}} e^{-i(k_i x_i -\d_i)}
\ri) r \hat k_j \s_j \s_k  \nn \\
&& + ~F~e^{i \frac{\hat q \cdot \vec r}{a\hbar}}
~\hat q_j \s_j \s_k  \Big ]  \Bigg ]
 ~.
\la{row2}
\eea
Note that the only difference between Eqs.(\ref{row1}) and (\ref{row2}) is
in the order of $\s_k$ and $\s_j$ in the last two terms in the RHS. Thus,
adding the two equations and using $\{\s_k,\s_j \}=0$, these terms simply drop
out. Next, dividing the rest by
$
f_{\bar k}(x_i,k_i,\d_i) \equiv
\prod_{i=1\neq k}^d
\le(
e^{ik_i x_i} + e^{-i(k_i x_i -\d_i)}
\ri)~,
$
where the subscript $\bar k$ of $f_{\bar k}$ signifies the lack of dependence
on $(x_k,k_k,\d_k)$, we get
\bea
&& e^{ik_k x_k} + e^{-i(k_k x_k -\d_k)}
+ f_{\bar k}^{-1} F~e^{i \frac{\hat q \cdot \vec r}{a\hbar} }
=
\nn \\
&&\hspace{-1cm}
\pm i
\le(
e^{ik_k x_k} - e^{-i(k_kx_k -\d_k)}\ri ) r \hat k_k
\pm i f_{\bar k}^{-1} F~e^{i \frac{\hat q \cdot \vec r}{a\hbar} } \hat q_k ~.
\la{bc5}
\eea
Note that for all practical purposes the boundary condition has factorized
into its Cartesian components, at least in the $a$ independent terms, which contain
$(x_k,k_k,\d_k)$ alone, i.e. no other index $i$.
Eq.(\ref{bc5}) yields, at $x_k=0$ and $x_k=L_k$, respectively,
\bea
e^{i\d_k} \le( 1 + i r \hat k_k \ri) = \le( i r \hat k_k -1 \ri)
+ f_{\bar k}^{-1} F'_k e^{-i\theta_k}
\la{bc7}
\eea
and
\bea
&& e^{i(2k_k L_k -\d_k)} \le( 1 + i r \hat k_k \ri) \nn \\
&& = \le( i r \hat k_k -1 \ri)
+ f_{\bar k}^{-1} F'_k e^{i\theta_k} e^{i\frac{|q_k|L_k}{a\hbar}}e^{i(k_k L_k -\d_k)}
\la{bc8}
\eea
where $F'_k \equiv \sqrt{1+ |\hat q_k|^2} F $, $\theta_k \equiv \arctan \hat q_k$
and we have assumed that $f_{\bar k}$ is evaluated at the same $x_i~(i\neq k)$ at
both boundaries of $x_k$.
Comparing Eqs.(\ref{bc7}) and (\ref{bc8}), which are the $d$-dimensional generalizations
of Eqs.(\ref{cond3}) and (\ref{cond4}), we see that
the following relations must hold
\bea
k_k L_k &=& \d_k = \arctan \le( - \frac{\hbar k_k}{mc} \ri) + {\cal O}(a)
\la{3dquant3} \\
\frac{|\hat q_k| L_k}{a \hbar} &=&
\frac{|\hat q_k | L_k}{a_0 \ell_{Pl}}
= 2p_k \pi - 2\theta_k ~.
\la{3dquant4}
\eea
While Eq.(\ref{3dquant3}) yields quantization of energy levels in $d$ dimensions
($k_k L_k = n\pi$ in the non-relativistic limit),
Eq.(\ref{3dquant4}) shows that lengths in all directions are quantized.
Further, one may choose the symmetric case
$|\hat q_k | = 1/\sqrt{d}$ \footnote{Alternatively,
assuming no direction is intrinsically preferred in space and
the only {\it special} direction is provided by the particle momentum $\vec k$,
one can make the identification $\hat q=\hat k$, in which case
$|\hat q_k |= n_k/\sqrt{ \sum_{i=1}^d n_i^2 } \approx 1/\sqrt{d}$, assuming
that the momentum quantum numbers $n_k \gg 1$ and approximately equal, when space is probed at the
fundamental level with ultra high energy super-Planckian particles.},
in which case, it follows from Eq.(\ref{3dquant4}) above
\bea
\frac{L_k}{a_0 \ell_{Pl}}
= \le( 2p_k \pi - 2\theta_k\ri) \sqrt{d}~,~~p_k \in \mathbb{N}
\eea
which reduces to Eq.(\ref{1dquant}) for $d=1$. Note that the above also
gives rise to quantization of measured areas ($N=2$) and volumes ($N=3$), as follows
\bea
A_N \equiv
\prod_{k=1}^N \frac{L_k}{a_0 \ell_{Pl}}
= d^{N/2} \prod_{k=1}^N \le( 2p_k \pi - 2\theta_k \ri)~,~~p_k \in \mathbb{N}~.\hspace{1.5ex}
\eea
%
%
\section{Spherical Cavity: Dirac Equation in Polar Coordinates}
%
%
%
Finally, we solve the Dirac equation with the GUP-induced terms in a spherical cavity, and show that
only cavities of certain discrete dimensions can confine a relativistic particle. We
follow the analysis of \cite{bhaduri}. For related references, see \cite{thomas,hey}.
A spherical cavity of radius $R$, defined by the potential
\bea
U(r) &=& 0, ~r \leq R~, \nn \\
       &=& U_0 \rightarrow \infty,~r>R
\eea
yields the following Dirac equation in component form
\bea
c \le(\vec\sigma \cdot \vec p_0 \ri)\chi_2 + \le( mc^2  + U \ri)\chi_1 - c a p_0^2 \chi_1  &=&  E\chi_1 \la{rkb1} \\
c \le(\vec\sigma \cdot \vec p_0 \ri)\chi_1 - \le( mc^2 + U \ri)\chi_2 - c a p_0^2 \chi_2  &=&  E\chi_2 \la{rkb2}
\eea
where the Dirac spinor has the form
$\psi = \left( \begin{array}{c}
\chi_1 \\
\chi_2 \end{array} \right)$~.
It can be shown that the following operators commute with the GUP-corrected Hamiltonian:
the total angular momentum operator (not to be confused with the Dirac current represented by the
same letter)
$
\vec J = \vec L + \vec\Sigma/2~,~K = \beta \le(\vec\Sigma \cdot \vec L + I \ri)~,
$
where
$
\vec L
$
is the orbital angular momentum,
$
\vec\Sigma =  \left( \begin{array}{cc}
\vec\sigma & 0 \\
0 & \vec\sigma
\end{array} \right)~,
$
and $K^2=J^2 + 1/4$.
Thus, eigenvalues of $J^2$ and $K$, namely $j(j+1)$ and $\kappa$ respectively, are related by
$\kappa = \pm(j+1/2)$. Correspondingly, the Dirac spinor has the following form
\bea
\psi &=& \left( \begin{array}{c}
\chi_1 \\
\chi_2 \end{array} \right)
= \left( \begin{array}{c}
g_\kappa(r) {\cal Y}^{j_3}_{j\ell}(\hat r)
 \\
if_\kappa(r) {\cal Y}^{j_3}_{j\ell'}(\hat r)
 \end{array} \right)~, \\
{\cal Y}^{j_3}_{j\ell} &=&
\le(
l~~\frac{1}{2}~~j_3-\frac{1}{2}~~\frac{1}{2}~
\rule[-0.22cm]{.01cm}{.7cm}
~j~~j_3\ri)~Y^{j_3-\frac{1}{2} }_{\ell} (\hat r)
\left(
\begin{array}{c}
1
\\
0
\end{array}
\right) \nn \\
&+&
\le(
l~~\frac{1}{2}~~j_3+\frac{1}{2}~~-\frac{1}{2}~
\rule[-0.22cm]{.01cm}{.7cm}
~j~~j_3\ri)~Y^{j_3+ \frac{1}{2} }_{\ell} (\hat r)
\left(
\begin{array}{c}
0
\\
1
\end{array}
\right)\hspace{4ex}
\eea
where $Y^{j\pm\frac{1}{2}}_l$ are spherical harmonics and
$\le(
j_1~j_2~m_1~m_2~
\rule[-0.2cm]{.01cm}{.6cm}
~J~M\ri)
$
are Clebsch-Gordon coefficients. $\chi_1$ and $\chi_2$ are eigenstates of $L^2$     with eigenvalues
$\hbar^2 \ell(\ell + 1)$ and $\hbar^2 \ell'(\ell'+ 1)$, respectively,  such that the following hold
\bea
\mbox{if} ~~\kappa &=& j+\frac{1}{2}>0~,\nn\\
\mbox{then} ~~ \ell&=&\kappa=j+\frac{1}{2},~~\ell'=\kappa-1=j-\frac{1}{2}~,  \\
\mbox{and}~\mbox{if} ~~\kappa &=& -(j+\frac{1}{2})<0~,\nn\\
\mbox{then} ~~\ell&=&-(\kappa+1)=j-\frac{1}{2},~\ell'=-\kappa=j+\frac{1}{2}~.\hspace{2ex}
\eea
Next, we use the following identities
\bea
%
%
\vec \sigma\cdot \vec p_0 &=& \frac{\vec\sigma\cdot \vec r}{r^2}
\le[(\vec \sigma\cdot \vec r)(\vec \sigma \cdot \vec p_0) \ri]
\nn\\
&=&
\frac{\vec\sigma\cdot \vec r}{r^2}
\le[
\vec r \cdot \vec p_0 + i \vec\sigma \cdot \vec r \times \vec p_0
\ri]
\nn\\
&=&
\frac{\vec\sigma\cdot \vec r}{r^2}
\le[
- i \hbar r \frac{d}{dr}
+ i \vec\sigma\cdot\vec L
\ri] \\
&& \le(\vec\sigma\cdot \vec L +1
\ri)\chi_{1,2} = \mp \kappa \chi_{1,2} \\
&&\le(
\vec\sigma \cdot \hat r
\ri) {\cal Y}^{j_3}_{j\ell} = - {\cal Y}^{j_3}_{jl'}~,~
\le(
\vec\sigma \cdot \hat r
\ri) {\cal Y}^{j_3}_{j\ell'} = - {\cal Y}^{j_3}_{jl}\hspace{3ex}
\eea
where we have used
$(\vec\sigma\cdot \vec A)(\vec\sigma \cdot \vec B) = \vec A \cdot \vec B + i \vec \sigma \cdot (\vec A \times \vec B)$,
the related identity
$(\vec\sigma\cdot \vec r)(\vec\sigma \cdot \vec r) =r^2$, and the relation
$
p_0^2 F(r) Y^m_\ell = \hbar^2
\le[
 - \frac{1}{r^2} \frac{d}{dr}
\le(
r^2 \frac{d}{dr} \ri)
+ \frac{\ell(\ell+1)}{r^2}
\ri] F(r) Y^m_\ell
$
for an arbitrary function $F(r)$, to obtain from Eqs.(\ref{rkb1})-(\ref{rkb2})
\bea
&&- c\hbar \frac{df_\kappa}{dr}+ c \frac{(\kappa-1)}{r} f_\kappa + (m c^2 +U)g_\kappa \nn  \\
&& + c a \hbar^2
\le[
\frac{1}{r^2} \frac{d}{dr}\le(
r^2 \frac{dg_\kappa}{dr} \ri) - \frac{\ell(\ell+1)}{r^2} g_\kappa
\ri]
= E g_\kappa \la{rkbgup1} \\
&& c\hbar \frac{dg_\kappa}{dr}+ c \frac{(\kappa+1)}{r} g_\kappa - (m c^2 +U)f_\kappa \nn  \\
&& + c a \hbar^2
\le[
\frac{1}{r^2} \frac{d}{dr}\le(
r^2 \frac{df_\kappa}{dr} \ri) - \frac{\ell'(\ell'+1)}{r^2} f_\kappa
\ri]
= E f_\kappa~.\hspace{2ex}\la{rkbgup2}
\eea
As in the case of rectangular cavities, Eqs.(\ref{rkbgup1})-(\ref{rkbgup2}) have the
standard set of solutions, slightly perturbed by the GUP-induced term (represented by
the ${\cal O}(a)$ terms below)
\be
g_\kappa(r)  = \tilde{N} j_\ell(p_0 r) + {\cal O}(a) ,
\ee
\be
~\mbox{where}~~\ell = \left\{ \begin{array}{l}
\kappa~, ~\mbox{if}~~\kappa>0 \\
-(\kappa+1), ~\mbox{if}~~\kappa<0\nn\\
\end{array} \right.
\ee
\be
f_\kappa(r)  = \tilde{N}  \frac{\kappa}{|\kappa|}
\sqrt{\frac{E-mc^2}{E+mc^2}}
 j_{\ell'}(p_0r)+ {\cal O}(a) ,
 \ee
\be
~\mbox{where}~~\ell' = \left\{ \begin{array}{l}
(\kappa-1)~, ~\mbox{if}~~\kappa>0 \\
-\kappa, ~\mbox{if}~~\kappa<0\nn\\
\end{array} \right.
\ee
where $j_l(x)$ are spherical Bessel functions.
It can be shown that the MIT bag boundary condition (at $r=R$) is equivalent to \cite{thomas,bhaduri}
\bea
\bar\psi_\kappa \psi_\kappa = 0
\la{mitbagbc2}
\eea
which in the massless (high energy) limit yields
\bea
\le[ g_\kappa^2(r) - f_\kappa^2(r) \ri] \le({\cal Y}^{j_3}_{jl}\ri)^\dagger
{\cal Y}^{j_3}_{jl} + {\cal O}(a)
=0 \la{rkbeigenvalue}
\eea
which in turn gives the quantization of energy (for energy eigenvalues obtained numerically from
Eq.(\ref{rkbeigenvalue}), see Table 2.1, Chapter 2, ref.\cite{bhaduri}. These will also undergo tiny
modifications ${\cal O}(a)$.).

But from the analysis of previous sections, we expect new non-perturbative solutions of the form
$f_\kappa = {\cal F}_\kappa(r) e^{i\epsilon r/a\hbar}$ and
$g_\kappa = {\cal G}_\kappa(r) e^{ i\epsilon r/a\hbar}$
(where $\epsilon = {\cal O}(1)$)
for which Eqs.(\ref{rkbgup1})-(\ref{rkbgup2}) simplify to
\bea
a \hbar \frac{d^{2}g_\kappa}{dr^2}&=& \frac{df_\kappa}{dr} \\
a \hbar \frac{d^{2}f_\kappa}{dr^2}  &=& - \frac{dg_\kappa}{dr}
\eea
where we have dropped terms which are ignorable for small $a$. These indeed have solutions
\bea
f_\kappa^{\cal N} &=& i N' e^{i r/a\hbar } \\
g_\kappa^{\cal N} &=& N' e^{ i r/a\hbar}~,
\eea
where similar to the constant $C$ in ref.\cite{advplb}, here one must have
$\lim_{a\rightarrow 0}N' =0$, such that these new solutions drop out in the
$a\rightarrow 0$ limit.
The boundary condition (\ref{mitbagbc2}) now gives
\bea
|g_\kappa(r) + g_\kappa^{\cal N}(r)|^2 = |f_\kappa(r) + f_\kappa^{\cal N}(r)|^2~,
\eea
which to ${\cal O}(a)$ translates to
\bea
&& \le[
j_\ell^2(p_{0} R) - j_{\ell'}^2(p_{0}R)
\ri] \nn\\
&& + 2N' \le[
j_\ell(p_{0}R)   \cos( R/a\hbar) - j_{\ell'}(p_{0}R) \sin (R/a\hbar)
\ri] = 0 ~.\hspace{5ex}
\eea
This again implies the following conditions
\bea
j_\ell (p_{0}R) &=&  j_{\ell'}(p_{0}R)  \\
\tan(R/a\hbar) &=& 1~.
\eea
The first condition is identical to Eq.(\ref{rkbeigenvalue}), and hence the energy quantization.
The second implies
\bea
\frac{R}{a\hbar} = \frac{R}{a_0 \ell_{Pl}}=  2p\pi - \frac{\pi}{4} ~,~~p \in \mathbb{N} ~.
\eea
This once again, the radius of the cavity, and hence the area and volume of spheres are seen to be quantized.
%
%
\section{Conclusions}
%
%
%
In this Letter, we have studied a relativistic particle in a box in one, two and three dimensions
(including a spherical cavity in three dimensions),
using the Klein-Gordon and Dirac equations with corrections that follow from the Generalized Uncertainty Principle.
We have shown that to confine the particle in the box, the dimensions of the latter would have to
be quantized in multiples of a fundamental length, which can be the Planck length.
As measurements of lengths, areas and volumes require the existence and use of such particles,
we interpret this as effective quantization of these quantities. Note that although existence of a
fundamental length is apparently inconsistent with special relativity and Lorentz transformations
(fundamental length in {\it whose frame?}),
it is indeed consistent, and in perfect agreement with Doubly Special Relativity Theories.
It is hoped that the essence of these results will continue to hold in curved spacetimes, and even if possible
fluctuations of the metric can be take into account in a consistent way. In addition to exploring these issues,
it would be interesting to study possible phenomenological implications of space quantization; e.g. if it has
any measurable effects at distance scales far greater than the Planck length, such as at about $10^{-4}~fm$, the
length scale to be probed by the Large Hadron Collider. We hope to make further studies in this direction
and report elsewhere.

%
%
%

%
%
\vs{.1cm}
\noindent\\
{\bf Acknowledgment}\\
We specially thank R. K. Bhaduri for discussions, useful correspondence and suggestions.
We thank A. Dasgupta, S. Ghosh, R. B. Mann, R. Parwani and L. Smolin for
discussions, P. Alberto for correspondence and the anonymous referee
for an important suggestion regarding particles with integer spins.
This work was supported in part by the Natural
Sciences and Engineering Research Council of Canada and by the
Perimeter Institute for Theoretical Physics.
%
%
%
%
%
\vs{.1cm}
\noindent\\
{\bf Note added}\\
%
%
%
Following a suggestion of the anonymous referee to re-examine wave equations for a
spinless particle with relativistic corrections, we explore two routes : \\
(i) First, we write the time-dependent version of the GUP-corrected Dirac equation (\ref{ham1})
\bea
H \psi &=& \le (c\, \vec \a \cdot \vec p + \b mc^2 \ri) \psi (\vec r,t)  \nn \\
&=&  \le(c\, \vec \a \cdot \vec p_0 -
c\, a (\vec\a \cdot \vec p_0)(\vec\a \cdot \vec p_0) + \b mc^2 \ri) \psi (\vec r,t)\nn \\
&=& i\hbar \frac{\pa \psi (\vec r,t)}{\pa t}~.
\la{ham2}
\eea
To study the non-relativistic limit, we write the spinor $\psi$ in a slightly different form as
\cite{mannbook}
\bea
\psi &=& e^{-\frac{i}{\hbar}mc^{2}t}\left( \begin{array}{c}
\chi_1 (\vec r,t) \\
\chi_2 (\vec r,t) \end{array} \right) ~,
\eea
and obtain the two component equations
\bea
i\hbar \frac{\pa \chi_1}{\pa t} &=&  c\le( \vec\sigma\cdot\vec p_0 \ri) \chi_2
-ca \le( \vec\sigma\cdot\vec p_0 \ri)^2 \chi_1
\la{chi1} \\
i\hbar \frac{\pa \chi_2}{\pa t} &=& -2mc^2 \chi_2  + c \le( \vec\sigma\cdot\vec p_0 \ri) \chi_1
-ca \le( \vec\sigma\cdot\vec p_0 \ri)^2 \chi_2 ~.\hspace{4ex}
\la{chi2}
\eea
For $mc^2 >> |\pa \chi_2/\pa t|$, Eq.(\ref{chi2}) gives to ${\cal O}(a)$
\bea
\chi_2
=
\frac{1}{2mc}
\le[1  - \frac{a}{2mc} \le( \vec\sigma\cdot\vec p_0 \ri)^2  \ri]
\le( \vec\sigma\cdot\vec p_0 \ri) \chi_1
~,
\eea
which when substituted in Eq.(\ref{chi1}) leads to
\bea
&& i\hbar \frac{\pa \chi_1}{\pa t} =
 \frac{1}{2m}\le( \vec\sigma\cdot\vec p_0 \ri)^2 \chi_1 \nn \\
&-&  \frac{a}{(2m)^2c} \le( \vec\sigma\cdot\vec p_0 \ri)^4
\chi_1 -  ca \le( \vec\sigma\cdot\vec p_0 \ri)^2 \chi_1~.
\la{chi2b}
\eea
Using the identity
$\le( \vec\sigma\cdot\vec p_0 \ri)^2=p_0^2$,
Eq.(\ref{chi2b}) becomes
\bea
 i\hbar \frac{\pa \chi_1}{\pa t} =
\le[
\le( \frac{1}{2m} -ca \ri) p_0^2
-\frac{a}{(2m)^2c} p_0^4
 \ri] \chi_1~.
\eea
Finally, substituting
$\chi_1 (\vec r,t) = e^{-iEt/\hbar}\chi_1(\vec r)$, we get
\bea
\le[
\le( \frac{1}{2m} -ca \ri) p_0^2
-\frac{a}{(2m)^2c} p_0^4
 \ri] \chi_1 = E\chi_1~.
\la{mann1}
\eea
%
%
Although the above GUP-corrected Pauli equation actually describes a $2$-component,
non-relativistic spinor, it is an interesting (and new)
extension of the Schr\"odinger equation, and can have potential applications elsewhere.
%
Note that the above holds in any spacetime dimension and is local.
In particular, in one dimension, in addition to the usual plane wave solutions
$\chi_1 = e^{\pm i k' z}$ (where $k'=\sqrt{2mE/\hbar^2} + {\cal O}(a)$), Eq.(\ref{mann1})
also admits of the non-perturbative solutions $\chi_1 = e^{\pm \frac{i}{\hbar} \sqrt{2mc/a}~z}$,
which too are plane waves but with wavelength $\approx \hbar\sqrt{a/mc}=\sqrt{a_0 \ell_{pl}\hbar/mc}$.
Imposing standard boundary condition $\chi_1=0$ at $z=0$ and $z=L$ and following
the procedure outlined in \cite{advplb}, it is easy to show that $L/\sqrt{a_0\ell_{Pl}}$ is quantized.

\noindent
(ii) Next, we square the operators on both sides of Eq.(\ref{ham1}), use $\b^2=1$ and the
relation $(\vec\a \cdot \vec p_0)^2=p_0^2$
to obtain
\bea
H^2 &=&
p_0^2 c^2 + m^2c^4 - 2ca p_0^2\le[c \vec\a\cdot\vec p + \b m c^{2} \ri] \nn \\
&=& p_0^2 c^2 + m^2c^4 - 2ca p_0^2 H~\nn \\
&=&
 p_0^2 c^2 (1-2aH/c) + m^2c^4 \la{ham3}
\eea
where in the last term of the intermediate step we have substituted $H=c \vec\a\cdot\vec p_0 + \b m c^{2} + {\cal O}(a)$.
It is seen that the above too can be used in any dimension and is local. Furthermore, by construction,
solutions of (\ref{ham1}) are also solutions of (\ref{ham3}) treated as a differential equation,
resulting in identical space quantization results.

We expect similar results to hold for
equations governing bosonic fields with higher spins, such as Maxwel's equations, including GUP corrections.
It would be interesting to see the interplay of such a field with fermions, say via minimal coupling.
We hope to report on it elsewhere.
%
%
%
%
%
%
%
%


\end{document}